\documentclass[epj]{svjour}
\usepackage{graphics}
\begin{document}
\title{Distance statistics in random media}
\subtitle{High dimension and/or high neighborhood order cases}
\author{Cristiano Roberto Fabri Granzotti\inst{1,}\thanks {\email{c.roberto.fg@usp.br}} \and Alexandre Souto Martinez\inst{1,2,}\thanks {\email{asmartinez@ffclrp.usp.br}}
}                     
%
%
\institute{
           	Faculdade de Filosofia, Ci\^encias e Letras de Ribeir\~ao Preto (FFCLRP),
           	Universidade de S\~ao Paulo (USP),
           	Avenida Bandeirantes, 3900,
           	14040-901, Ribeir\~ao Preto, S\~ao Paulo, Brazil\and
           	National Institute of Science and Technology in Complex Systems (LNCT-SC),
           	Universidade de S\~ao Paulo (USP),
           	Avenida Bandeirantes, 3900,
           	14040-901, Ribeir\~ao Preto, S\~ao Paulo, Brazil        
           }

\date{Received: date / Revised version: date}
%
\abstract{
Consider an unlimited homogeneous medium disturbed by points generated via Poisson process. 
The neighborhood of a point plays an important role in spatial statistics problems. 
Here, we obtain analytically the distance statistics to $k$th nearest neighbor in a $d$-dimensional media. 
Next, we focus our attention in high dimensionality and high neighborhood order limits. 
High dimensionality makes distance distribution behavior as a delta sequence, with mean value equal to Cerf's conjecture. 
Distance statistics in high neighborhood order converges to a Gaussian distribution. 
The general distance statistics can be applied to detect departures from Poissonian point distribution hypotheses as proposed by Thompson and generalized here. 
} 
\maketitle
\section{Introduction}
\label{intro}
Consider a $d$-dimensional, unbounded, isotropic and homogeneous medium with disorder (points) generated by a Poisson process. The expected number of points in a 
volume $V_{d}$ is $\lambda = \rho V_{d}$, where $\rho$ is the point density. 
This disordered medium, although unlimited, can be represented
computationally as a $d$-dimensional hypercube, containing $N$ 
coordinates randomly distributed with uniform probability density function (pdf) along each edge (random point problem~\cite{BJP_Tercariol_2006_1}). This is a possible way to construct a disordered medium, where, distances among the points are not fixed, but vary statistically. In this medium, it is possible to exploit the neighborhood and distance statistics. 

Neighborhood statistics quantifies the probability of a point to be the $m$th nearest neighbor of its $n$th nearest neighbor. 
For $N \gg 1$, this probability was firstly calculated by Clark and Evans~\cite{Science_Clark_1955_1}, for $m=n=1$ and later generalized by Clark for mutual neighbors $m=n$ ~\cite{Science_Clark_1956_1}. 
Dacey corrected expression obtained by Clark~\cite{GA_Dacey_1969_1}. 
Neighborhood statistics was generalized by Cox, for $m \neq n$~\cite{IBT_Cox_1981_1} and interpreted in terms of multinomial distribution by Ter\c{c}ariol {\it et al.}~\cite{JPA_Tercariol_2007_1}.

Distance statistics quantifies the distance distribution of a given point to its $k$th nearest neighbor and can be applied in several disciplines. 
In Physics and Biology, this statistics can be used, for instance, in calculating the average separation between stars~\cite{RMP_Chandrasekhar_1943_1}, aggregation in plant community~\cite{E_Clark_1954,E_Thompson_1956}, optimal tour for Euclidean salesman problem~\cite{JPI_Cerf_1997_1,IJMP_Anirban_2001_1}, Euclidean matching problem~\cite{El_Mezard_1988,EPJB_Houdayer_1998}, partially self-avoiding walks~\cite{PRE_Lima_2001_1,PRE_Tercariol_2007_1}, thin films~\cite{AM_Tewari_2006_1}, etc. 
In Computer Science, the nearest neighbor distance statistics can be employed as pattern classifier~\cite{IT_Cover_1967,PR_Sameer_2001}, other than being used to determine distance between network terminals~\cite{IT_Martin_2005}, etc.

Up to now, only two aspects of this problem has been thoroughly addressed. 
The first one is the distance among points~\cite{AAM_Percus_1998_1,EJP_Pratip_2008_1} and the moments of the highest order~\cite{PRS_Evans_2002_1,PRS_Evans_2008_1} for different point distributions. 
The second one, is the distribution calculation for low-dimensional media, $d \le 3$, for nearest neighbor~\cite{E_Clark_1954,AM_Tewari_2006_1} and arbitrary neighborhood~\cite{E_Thompson_1956}. The distance distribution to the $k$th nearest neighbor in an arbitrary dimension has been calculated by Martin~\cite{IT_Martin_2005}, in the context of distance between internet access terminals. 
Despite the mathematical expression knowledge~\cite{IT_Martin_2005,AHN_Moltchanov_2012_1}, the parameters influence on the distribution have not been fully addressed, mainly in cases of high dimension and neighborhood order.

These limiting cases are non-trivial, because of the function ratio $\Gamma(z+x)/\Gamma(z)$, for $z \gg x$, where $\Gamma(z)$ is the gamma function. If one considers simpler expansion, inconsistencies like undefined central moments, such variance and skewness, occur. Our main contribution is to correct these inconsistencies using higher order terms in this ratio expansion. We calculate the distribution in these limiting cases and proof important results as the ones conjectured by Cerf {\it et al.}~\cite{JPI_Cerf_1997_1}.

In this paper, we obtain the analytical expressions for high dimensionality distance statistics that leads to an equivalence of the random link model. 
Also, the high neighborhood order is addressed, the resultant distribution converges to a Gaussian due to central limit theorem. 
The distance statistics can be used not only for predicting separation between neighbors, but also to detect departures from Poissonian hypothesis, as proposed by Thompson~\cite{E_Thompson_1956} and generalized here. 
Furthermore, we expand special cases of distance statistics varying dimensionality and neighborhood order.  

Our paper is organized as follows. In Sec. 2, we obtain the pdf of distance statistics in two distinct ways: 
throughout geometric interpretations and cumulative functions. This pdf is described by the generalized gamma distribution~\cite{AMS_Statcy_1962_1,Amoroso}. In Sec. 3, we calculate the high neighborhood order $k\gg 1$, and high dimension $d \gg 1 $ limiting cases. 
In this way, we demonstrated mathematically the Cerf's {\it et al.} conjecture~\cite{JPI_Cerf_1997_1}, and consider the combination of the limiting cases. 
In Sec. 4, we explore special cases of distance statistics by varying dimension and neighborhood order and propose a generalized hypothesis test to quantify deviations from Poissonian spatial process. Finally, on Sec. 5, we present the conclusions.
\section{Statistics of the random point problem}
\label{sec:1}

In this section, we obtain the analytical distance distribution expression and validate it by Monte Carlo simulations. In addition, we collapse the data with nearest neighbor distance distribution.
Consider a $d$-dimensional medium, with density $\rho$, where $\rho = \rho_{1}^{d}$ and $\rho_{1}$ is the one-dimensional medium density. 
The previous argument keeps the mean distance among points constant, which allow us to compare different system dimensionalities. 
The expected number of points in a hypersphere of radius $l$ is $\lambda = N_{u} l^d$, where $N_{u} = \rho \pi^{d/2}/\Gamma(1+d/2)$ is the number of points in a $d$-dimensional sphere with unitary radius. The probability of $k$ points to fall into a sphere of radius $l$ is given by the Poisson formula, $P(k) = e^{-\lambda}\lambda^{k}/k!$. 

The first method to derive the distance statistics is based on geometric arguments. 
The probability of $k$ points to fall inside a sphere of radius $l + dl$ is written as product of the probability of a sphere of radius $l$ to contain $k-1$ points and the probability of the spherical shell thickness $dl$ to contain only one point: $f_{\rho,d}^{(k)}(l)dl = P(k-1)P(1)$. As $dl \ll l$, the probability density function becomes:
\begin{equation}
f_{\rho,d}^{(k)}(l) = 
\frac{dN_{u}^k l^{dk-1}}{\Gamma(k)}
\exp(-N_{u}l^d),
\label{pdf}
\end{equation}
where $k$ is neighborhood order and can be mapped on the generalized gamma distribution: 
\begin{equation}
\mbox{GG}(x|\theta,k,\beta) =
\frac{1}{\beta\theta\Gamma(k)}
\left(\frac{x}{\theta}\right)^{k/\beta-1}
\exp\left[-\left(\frac{x}{\theta}\right)^{1/\beta}\right],
\label{distStacy}
\end{equation}
with: $ \beta = 1/d$ and $\theta = N_{u}^{-\beta} = [\rho \pi^{d/2}/\Gamma(1+d/2)]^{-\beta}$, which depends on the point density and medium dimension. 
It is non-trivially affected by medium symmetry. 
If one considers a computer simulations, $\theta$ is only affected by media boundaries through out $\rho$. 
If, in one hand, one considers a $d$-dimensional hypercube with edge length $\cal{L}$ and $N$ points, then $\rho = N/{\cal{L}}^{d}$. In the other hand, if one considers a sphere: $\rho = N\Gamma(1+d/2)/\pi^{d/2}{\cal{L}}^d$.

\begin{figure}[htpb]
\resizebox{0.50\textwidth}{!}{
  \includegraphics{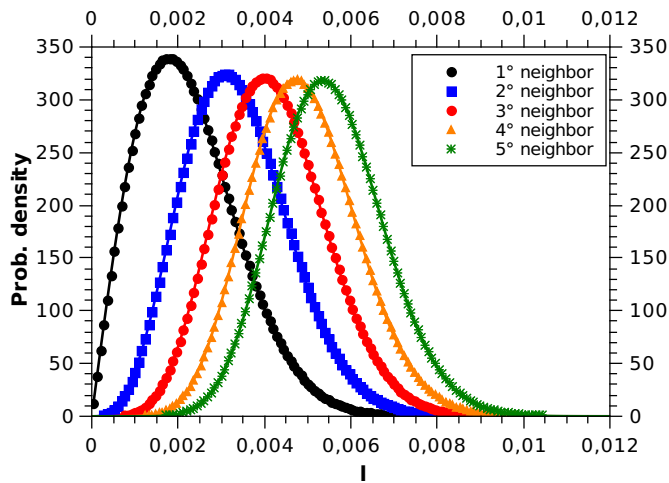}
}
\caption{Comparison of the analytical results generated by 
Eq.~\ref{pdf} (full lines) up to the fifth neighbor in a two dimension medium with Monte Carlo simulation. Simulations parameters are $\rho=50000$, with periodic boundary conditions. The increase of neighbor order makes distance statistics more symmetric with respect to the mean value.}
\label{neig5}
\end{figure}

Monte Carlo simulations validated Eq.~\ref{pdf}. 
The medium consisted of a cube with $N$ points and density $\rho = N/{\cal{L}}^{d}$. The results of Eq.~\ref{pdf} applied to this limited  medium is an approximation due to boundary effect, since points near the boundaries have fewer neighbors. 
Periodic boundary conditions minimize this effect. 
This validation is depicted in Fig.~\ref{neig5}, where one sees that increasing the neighborhood order, statistics distribution become more symmetric around their mean value. 
Moreover, the numerical experiments consider a finite number of points. The correction for finite size system is of order $1/N$ for the mean distance~\cite{AAM_Percus_1998_1}.
Further, Eq.~\ref{pdf} in terms of $\lambda = N_{u}l^{d}$, number of points in $d$-dimensional sphere of radius $l$, is collapsed into $f^{(k)}(\lambda) = 
\lambda^{k-1}e^{-\lambda}/\Gamma(k)$.

The second method is based on the cumulative distribution function. 
Consider firstly a point $i$ and its nearest neighbor, in two-dimensional medium. 
The probability of not finding any other point closer than $l$ is $P(k=0) = e^{-\rho \pi l^{2}}$. 
The random variable $L$, that describes the distance up to point $i$ nearest neighbor, $L > l$ if 
there is no points in area $\pi l^2$, consequently $P(L>l) = e^{-\rho \pi l^{2}}$. 
Thus, $L$ cumulative distribution function is: $P(L \leq l) = 1 - P(L > l) = F_{\rho,d}^{(1)}(l)$. 
The pdf that describes the distance to first neighbor is the $dF_{\rho,d}^{(1)}(l)/dl$: $f_{\rho,d}^{(1)}(l) = 2\rho\pi l e^{-\rho\pi l^{2}}$. This reasoning can be extended to arbitrary neighborhood and dimensionality, and leads to Eq.~\ref{pdf}. 

The mean distance of a point $i$ to its $k$th nearest neighbor is:
\begin{equation}
\langle l_{\rho,d}^{(k)}\rangle =
N_{u}^{-\beta}
\frac{\Gamma (k+\beta)}{\Gamma(k)}
\label{meandistance}
\end{equation}
which has been firstly obtained by Percus e Martin~\cite{AAM_Percus_1998_1}, 
and factorizes in density and neighborhood order. 
The $l_{\rho,d}^{(k)}$ variance is:
\begin{equation}
\sigma^{2}(l)_{\rho,d,k} = 
N_{u}^{-2\beta}
\left[
\frac{\Gamma(k+2\beta)}{\Gamma(k)} - 
\left(\frac{\Gamma(k+\beta)}{\Gamma(k)}\right)^2
\right],
\label{varmeandiastance}
\end{equation}
which is difficult to analyze, because of the $\beta=1/d$ ratio in the gamma function argument. 
When $k \gg \beta$, one can consider a simple expansion of ratio $\Gamma(k+\beta)/\Gamma(k)$ as $k^{\beta}$ or $k^{\beta}e^{-\beta/k}$, however the calculation of central moments as variance and skewness are inconsistent.
Expanding to higher orders, for $z\gg x$, one has:
\begin{equation}
\frac{\Gamma(z+x)}{\Gamma(z)} 
\approx 
z^{x}
\exp
\left(
\frac{-x}{2z}+
\frac{3x^{2}}{4z}
\right).
\label{gammaRatio}
\end{equation}

According to Eq.~\ref{gammaRatio}, the mean and standart deviation of $l_{\rho,d}^{(k)}$ can be approximated to: $\langle l_{\rho,d}^{(k)}\rangle \approx N_{u}^{-\beta}k^{\beta}$ and $\sigma(l)_{\rho,d,k} \approx c\beta N_{u}^{-\beta}k^{\beta-1/2}$, with $c=3/2$, 
indicating that the mean distance, in high dimensionality, is weakly affected by the neighborhood order, while the variance decays very rapidly. 
This occurs because the volume of a sphere is almost concentrated in a very thin spherical shell, when $d \gg 1$.
The skewness of Eq.~\ref{pdf} depends non-trivially on $k$ and $\beta$ 
\begin{equation}
\gamma_{1} = \frac{2-\Omega_{1}^{2}(k,\beta)/\Omega_{2}(k,\beta)+\Omega_{1}^{3}(k,\beta)/\Omega_{3}(k,\beta)}
{(1 -\Omega_{1}^{2}(k,\beta)/\Omega_{2}(k,\beta))^{3/2}},
\label{skewness}
\end{equation}
where $\Omega_{n}(k,\beta)=B(k,n\beta)/\Gamma(n\beta)$ and $B(a,b)= \Gamma(a)\\\Gamma(b)/\Gamma(a+b)$ is beta function. 
The skewness is modified only by neighborhood order and dimension, being independent of medium boundaries and density, using Eq.~\ref{gammaRatio} it can be approximated to $\gamma_{1} \approx 6\beta k^{-1/2}$. 
This simplification accurately describes the skewness behavior around the mean value due to the neighborhood order, see Fig.~\ref{neig5}. 
Also, the skewness factorizes in neighborhood order and dimensionality.

\section{Limiting cases}
\label{sec:2}

In this section, we analyze the behavior of Eq.~\ref{pdf}, firstly in the limit $d \gg 1$, next for $k \gg 1$ and finally both limits simultaneously. Although straightforward, these calculations present some pitfalls which are properly stressed.

\subsection{High dimensionality}

Let us introduce a new variable $y = (l- \langle l_{\rho,d}^{(1)}\rangle)/\sigma_{\rho,d,1}$, which standardizes the distance by the mean separation between the points. As $d \gg 1$, one has $\langle l_{\rho,d}^{(1)}\rangle \approx N_{u}^{-\beta}$ and $\sigma_{\rho,d,1}(l) \approx c\beta N_{u}^{-\beta}$. 
Distance can be rewritten as follows
\begin{equation}
l= 
N_{u}^{-\beta}
(1 + \beta cy).
\label{l_dinf}
\end{equation}
with $c=3/2$ and $\beta = 1/d$. In the $y$ variable, the pdf is obtained from the application of probability transformation law to Eq.~\ref{pdf}, using $l$ from Eq.~\ref{l_dinf}.
Starting with $k = 1$, one finds the Gumbel distribution ($1/|\bar{\lambda}|\exp[-(x/\bar{\lambda})-\exp(-x/\bar{\lambda})]$), with $\bar{\lambda} = -1/c$:
\begin{equation}
g(y) = c\exp[ cy - \exp(cy)],
\label{dinf_k1}
\end{equation}
which describes the minimum deviation from the expected separation: $\langle l_{\rho,d}^{(1)}\rangle = N_{u}^{-\beta}$. 
For higher neighborhood orders, one has:
\begin{equation}
g^{(k)}(y) = 
\frac{c}{\Gamma(k)}
\exp[cky - \exp(cy)],
\label{dinf_k}
\end{equation}
which is the log-gamma distribution ($[1/|\bar{\lambda}|\Gamma(k)]\exp[kx/\bar{\lambda} - \exp(x/\bar{\lambda})]$). 
The mean distance among points is calculated in two parts:
$ \langle y \rangle = \Psi(k)/c = (1/c)d[\ln\Gamma(k)]/dk$, that is digamma function~\cite{Book_Abramowitz_1972}, so that 
the mean distance among points is $\langle l_{\rho,d}^{(k)} \rangle = N_{u}^{-\beta} [1 + \beta\Psi(k)]$. 
The neighborhood order is an integer, which leads to a representation $\Psi(k) = -\gamma + \sum\limits_{i = 1}^{k - 1} i^{-1}$, rewriting $i^{-1}$ as $(k-i)^{-1}$, one has the mean distance on $l$: 
\begin{equation}
\langle l_{\rho,d}^{(k)} \rangle =
N_{u}^{-\beta}
\left[
1  +
\beta
\left(- \gamma + 
\sum\limits_{i = 1}^{k - 1}
\frac{1}{k-i}
\right)
\right]
\label{lmean_dinf},
\end{equation}
where $\gamma = 0.57721\ldots$ is Euler's constant and for $k \gg 1$, $\Psi(k) \approx \ln(k)$ and $\langle l_{\rho,d}^{(k)} \rangle = N_{u}^{-\beta}[1+\beta\ln(k)]$, this was firstly obtained by Cerf {\it et al.} conjecture expanding the term $\Gamma(k+\beta)/\Gamma(k)$ of Eq.~\ref{meandistance}. This term, on average, represent distance increment due to neighborhood order increase.
Due to accurate approximation for ratio $\Gamma(z+x)/\Gamma(z)$, we demonstrate this conjecture using distance statistics. 
Furthermore, Eq.~\ref{dinf_k} allows us to calculate not only mean distance, but also variance and higher order moments.
  
The variance of Eq.~\ref{dinf_k} is $\sigma^{2}(y)_{k} = \Psi^{(1)}(k)$, where $\Psi^{(1)}(k)$ is trigamma function. 
One can argue that $\sigma^{2}(a+bx) = b^{2}\sigma^{2}(x)$, so that the standard deviation in $l$ is
\begin{equation}
\sigma(l)_{\rho,d,k} = 
\frac{\beta N_{u}^{-\beta}}{\sqrt{k}},
\label{k_var_dinf}
\end{equation}
where we made use of the $\Psi^{(1)}(k) \approx 1/k$, for $k \gg 1$. 
In the $l$ variable, the mean distance is only weakly affected by neighborhood order, and the variance vanishes rapidly. This occurs because, in high dimensionality, a small increase in the radius leads to a large increase in volume. 
The larger the radius is, the smaller is the increment to generate the same increase in volume. 
Therefore, the higher the neighborhood order, the smaller the radius increase is and the lower the standard deviation is around the mean value. 
The distance distribution in variable $l$ is described as a delta sequence.
\subsection{High neighborhood order}

\begin{figure}[htpb]
\resizebox{0.50\textwidth}{!}{
  \includegraphics{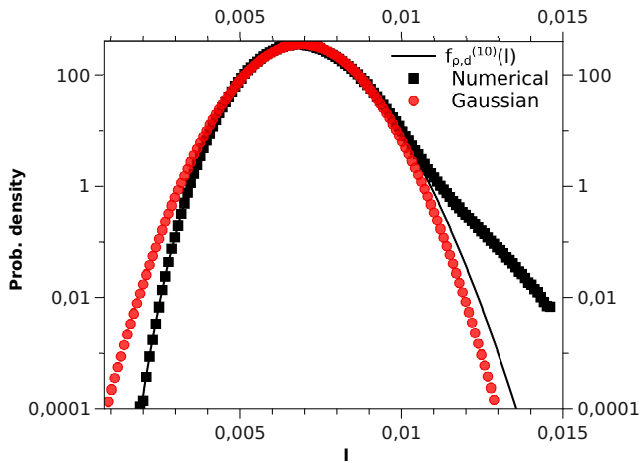}
}
\caption{Gaussian approximation for distance statistics for $k \gg 1$ and two dimensional media. Simulation parameters are $d=2$, $\rho = 65365$ and $k=10$. The weak adjustment  on the tails is due the fact that the ends of the distribution converge more slowly than the peak.}
\label{k_large}
\end{figure}

The second limiting case is the distance distribution for high neighborhood order. From Eq.~\ref{varmeandiastance}, one sees that the standard deviation decays according to $k^{\beta-1/2}$, for $k \gg \beta$. 
According to the central limit theorem, the variance of the summation $S$, of $N$ independent and identically distributed random variables, with finite variance, $\sigma$, is $\sigma_{r} = \sigma/ \langle S \rangle$ and the skewness decreases as $1/\sqrt{N}$~\cite{EPJB_Houdayer_1998}. 
For $k \gg 1$ and for arbitrary dimension, the relative standard deviation and skewness decrease as $1/\sqrt{k}$. This indicates that, besides recovering the symmetry of the pdf around its mean value, the neighborhood order increase makes Eq.~\ref{pdf} to converge to a Gaussian distribution. This behavior is obtained by numerical simulation and illustrated in the graphs of Figs.~\ref{neig5} and~\ref{k_large}. 

The convergence to the Gaussian is due to the summation of volumes. The necessary volume around a point $i$ to find $k$ neighbors is on average $kV_{1}$, where $V_{1}$ is the volume needed to find the nearest neighbor, distance in this case is a random variable proportional to $(kV_{1})^{\beta}$. 
Another way of understanding this convergence is considering the summation of the thicknesses of spherical shells comprising the same volume.
\section{Applications}
\label{sec:3}

In this section, we discuss possible applications of our results in context of pseudo number generation and tests of departures from Poissonian hypothesis in spatial distributions. Table(\ref{tab:1}) enumerates various pdfs obtained by varying dimensionality of the medium and neighbor order in Eq.~\ref{pdf}.

Due to the great amount of special cases of distance statistics, Table(\ref{tab:1}), one possible application is to use it as a very general pseudo random numbers generator. Although not efficient in terms of time consumption, it allows one to have a unified probability density functions arising from distance measurements in random media.

\begin{table}[htpb]
\centering
\caption{Summary of probability distributions obtained from Eq.~\ref{pdf} for different dimensionalities and neighborhood orders. The symbol (-) means arbitrary value, $\infty$ is a high value and (*) means distribution in $y$ variable given by Eq~\ref{l_dinf}.}
\label{tab:1}       
\begin{tabular}{lll}
\hline\noalign{\smallskip}
$d$ & $k$ & Distribution\\
\noalign{\smallskip}\hline\noalign{\smallskip}
1 & 1 & Exponential \\ 
1 & - & Gamma \\ 
1 & $\infty$ & Gaussian \\
2 & 1 & Rayleigh \\
2 & - & Nakagami \\
3 & - & Wilson-Hilferty \\
- & 1 & Weibull \\ 
- & - & Stacy \\
- & $\infty$ & Gaussian \\
$\infty$ & 1 & *Gumbel \\ 
$\infty$ & - & *Log-Gama \\ 
$\infty$ & $\infty$ & *Gaussian \\   
\noalign{\smallskip}\hline
\end{tabular}
\end{table}

Another possible application of Eq.~\ref{pdf} is to evaluate, whether given distances among points vary from Poissonian hypothesis. This evaluation was originally employed by Thompson~\cite{E_Thompson_1956}, on the distance distribution among trees in a two dimensional environment. 
One way to assess deviations from this hypothesis is to perform a test of significance for the average distance to the $k$th nearest neighbor. The test uses limits of Eq.~\ref{pdf}, when it is transformed into a $\chi^{2}$ (chi square) distribution, Eq.~\ref{departures}. 
As a generalization of Thompson result, we propose the same test in an environment of arbitrary dimensionality. 
Rewriting Eq.~\ref{pdf} as a function of $x_{n} = 2\lambda$, it becomes:
\begin{equation}
 f^{(k)}(x_{n}) = 
 \frac{1}{2\Gamma(k)}
 \left(
 \frac{x_{n}}{2}
 \right)^{k-1}
 \exp(-x_{n}/2),
 \label{departures}
\end{equation}
that is $\chi^{2}$ distribution, with $2k$ degrees of freedom. Once one knows the density of points on media, it is possible to apply the test and detect deviations in any neighborhood order, not only for the nearest one.
\section{Conclusion}
\label{conclusion}
Using only Poisson process, we calculate the statistical distribution of distance for disordered media with arbitrary dimensionality. 
Our results have been validated by Monte Carlo simulations. Starting with Eq.~\ref{pdf} we calculate the limiting case of high dimensionality and high neighborhood order. 
Distance statistics on high dimensional case becomes a delta sequence around the mean distance, that was firstly conjectured by Cerf \textit{et al.}.
Distance statistics in high neighborhood order converges to a Gaussian distribution due to central limit theorem. 
The general pdf with $d<3$ and arbitrary neighborhood order leads to special cases that retrieves well known pdfs such gamma, Weibull, etc. 
Distance statistics may detect departures from Poissoanian, as pointed by Thomson for $d=2$, and generalized by Eq.~\ref{departures}, opening up new possibilities like three dimensional image analyzes of cells distribution, etc.  

\begin{acknowledgement}
C.R.F.Granzotti acknowledges support from CAPES and FAP\\ESP(2010/00087-0).
A.S.M. acknowledges support from CNP\\q(305738/2010-0 and 485155/2013-3) and CAPES. The authors would like to thank C.A.S Terçariol, R. S. Gonz\'alez and J. M. Berbert for useful discussions.
\end{acknowledgement}

\end{document}